\documentclass{article}
\usepackage{cite}
\usepackage{graphicx}
\usepackage{hyperref}
\usepackage{epstopdf}
\usepackage[utf8]{inputenc}
\usepackage{amsmath,amssymb}
\usepackage{subcaption}
\usepackage{multirow}

\begin{document}

\title{Adaptive Event Sensing in Networks of Autonomous Mobile Agents}

\author{Rodrigo~R.~Esch$^1$\\
F\'abio~Protti$^2$\\
Valmir~C.~Barbosa$^1$\thanks{Corresponding author (valmir@cos.ufrj.br).}\\
\\
$^1$Programa de Engenharia de Sistemas e Computa\c c\~ao, COPPE\\
Universidade Federal do Rio de Janeiro\\
Caixa Postal 68511\\
21941-972 Rio de Janeiro - RJ, Brazil\\
\\
$^2$Instituto de Computaç\~ao\\
Universidade Federal Fluminense\\
Rua Passo da P\'atria, 156\\
24210-240 Niter\'oi - RJ, Brazil}

\date{}

\maketitle

\begin{abstract}
Given a connected region in two-dimensional space where events of a certain kind
occur according to a certain time-varying density, we consider the problem of
setting up a network of autonomous mobile agents to detect the occurrence of
those events and possibly record them in as effective a manner as possible. We
assume that agents can communicate with one another wirelessly within a fixed
communication radius, and moreover that initially no agent has any information
regarding the event density. We introduce a new distributed algorithm for agent
control based on the notion of an execution mode, which essentially lets each
agent roam the target region either at random or following its local view of a
density-dependent gradient. Agents can switch back and forth between the two
modes, and the precise manner of such changes depends on the setting of various
parameters that can be adjusted as a function of the application at hand. We
provide simulation results on some synthetic applications especially designed to
highlight the algorithm's behavior relative to the possible execution modes.

\end{abstract}

\newpage
\section{Introduction}\label{sec:intro}

There exist several applications in which sensing a geographical region as
accurately as possible can be crucial. Examples include search and recovery
operations, manipulation tasks in hazardous environments, surveillance, and
environmental monitoring for pollution detection \cite{Gusrialdi2011}. Often it
is the case that monitoring the region in question with only one sensor to
detect events all across it can be extremely expensive if not downright
infeasible, as well as hardly fault-tolerant or scalable. In recent years,
though, significant technological advances, especially those making
communication, processing, and sensing more cost-effective, have enabled the use
of autonomous mobile agents \cite{Martinez2007}.

If used in sufficient numbers given the size of the region to be monitored,
these agents offer a viable alternative to overcome the limitations of the
single-sensor scenario, provided only that they can perform local sensing and
communicate with one another. Once put to work together on the same global
sensing task, the agents can communicate to one another the sensing results
obtained with their own relatively small sensing capabilities, thus generating
a far greater sensing power for the detection of the majority of events
occurring in the region. The key to realize such a tantalizing possibility lies,
naturally, in how successfully the agents can cooperate toward a common goal.
 
One key additional ingredient is then agent mobility. Mobile agents can in
principle behave in several different modes, such as moving to places where no
events are currently occurring so that they can be detected when they do occur;
moving to places where a great number of events is currently occurring and
additional sensing power is needed; or simply moving in arbitrary directions
aiming to discover unanticipated places where events of interest may come to
occur or be already occurring. Given a specific sensing task, a successful
ensemble of agents will be one whose agents behave in the appropriate manner
often enough for the task to be accomplished accurately and with just enough
agents.

Here we introduce a new distributed algorithm for the operation of such a group
of autonomous mobile agents. Our algorithm is based on the notion of an
execution mode to guide each agent toward participating in the overall
sensing task as effectively as possible while switching back and forth from one
mode to another as mandated by local circumstances and parameter values. As in
so many cases in which agent adaptation has a key role to play, here too we aim
to strike a fruitful balance between exploration and exploitation. We do so by
using two execution modes exclusively. The first one, called the random mode,
is designed so that the agent can contribute to the overall task at hand by
exploring new territory that so far may have remained insufficiently monitored.
The second mode, referred to as the gradient mode, lets the agent exploit the
best sensing opportunities available to it by focusing on those places at which
events are deemed more likely to occur insofar as the agent's local view of the
global event density allows such a conclusion.

Our work is preceded by important related research providing both improvement
opportunities and inspiration
\cite{
Howard2002,
Zou2003,
Li2005,
Nene2010,
Gusrialdi2011,
Gusrialdi2013,
Song2013}.
Some of this research has addressed the problem of maximizing coverage area when
the agents are placed inside the target region either deterministically
\cite{Howard2002} or at random \cite{Zou2003,Nene2010}, and also the similar
problem of adjusting the agents' initial locations and their speeds along a
closed path so that all points inside such a perimeter are monitored as fully
as possible \cite{Song2013}. Another related problem addressed by such research
has been the convergence of all sensed data onto a base station while being
mindful of the energy spent on communication \cite{Li2005}.

Although all these problems are undoubtedly related to the one we tackle, the
coverage-area maximization that lies at their core is only ancillary to us, in
the sense that what we seek is full event coverage even if this means leaving
some portions of the target region somewhat unmonitored. We share this goal with
the works in \cite{Gusrialdi2011,Gusrialdi2013}, but solve the more general
problem in which event density is both time-varying and unknown to the agents
initially. Moreover, the problem is to be solved subject to the further
constraints that agents need not be within direct communication reach of one
another, and that no centralized element (like the leader-following procedure
from \cite{Meng2007} that the authors of \cite{Gusrialdi2011,Gusrialdi2013}
adopt) is to be part of the solution.

We proceed in the following manner. First we lay down our assumptions regarding
both the events to be monitored and the agents that will monitor them. We do
this in Section~\ref{sec:theproblem}. Then we move to a detailed description of
our solution in Section~\ref{sec:solution}. Simulation results on a few key
scenarios are given in Section~\ref{sec:experiments}, which is then followed by
conclusions in Section~\ref{sec:conclusion}.

\section{System model}\label{sec:theproblem}

The target region is modeled as a connected set $\Omega \in \mathbb{R}^2$. In
this region, an event happens at the infinitesimal vicinity $\mathrm{d}q$ of
point $q \in \Omega$ with probability proportional to $\phi(q, t) \mathrm{d}q$,
where $t$ is continuous time and $\phi(q, t)$ is a time-dependent event density
function such that $\int_{\Omega} \phi(q, t) \mathrm{d}q < \infty$ for all
$t \geq 0$. The function $\phi$ is unknown to all agents initially.

We assume that the occurrence of an event leaves a footprint that only
disappears after a number of time units given by $\mathit{VisTime}$, a
parameter, have elapsed. Upon coming across such a footprint, an agent is
capable of estimating the corresponding event's time of occurrence. A useful
example here is the indirect detection of a past fire at a certain point in
$\Omega$ by means of the temperature at that point when it is reached by an
agent. We refer to $\mathit{VisTime}$ as an event's visibility time and assume
it is the same for all events, given a specific domain of interest.

The probability that agent $i$, located at point $s_i \in \Omega$, detects an
event (or its footprint) occurring at point $q \in \Omega$ is modeled as
\begin{equation}
	\label{eq:eq-1}
	p_i(q) = \left\{ 
	\begin{array}{l l}
		\left(1- \frac{d_i}{R_s}\right)^2 & \quad \text{if $d_i \le R_s$}\\
		0 & \quad \text{otherwise,}\\
	\end{array} \right.
\end{equation}
where $d_i$ is the Euclidean distance between points $s_i$ and $q$ and $R_s$ is
the sensor's maximum sensing range. That is, we assume that the agent's sensing
capability decays with the distance to the point of occurrence as a convex
parabola, provided this point is located within the circle of radius $R_s$
centered at the agent's location. No sensing is possible outside this circle.
Whenever sensing is possible, we assume it to be instantaneous.

Agents can communicate with one another wirelessly. When agent $i$ is at point
$s_i$, any message it sends is assumed to be instantaneously received and
successfully decoded by some other agent $j$, located at point $s_j$, with
probability
\begin{equation}
	\label{eq:eq-2}
	r_i(s_j) = \left\{ 
	\begin{array}{l l}
		1 & \quad \text{if $d_{ij} \le R_c$}\\
		0 & \quad \text{otherwise,}\\
	\end{array} \right.
\end{equation}
where $d_{ij}$ is the Euclidean distance between agents $i$ and $j$ and $R_c$ is
the maximum separation between two agents across which they can still
communicate. We remark that, should an undirected graph be used to model the
communication possibilities among all agents at a certain point in time, such a
graph would follow what is known as the Boolean model. This model is widely used
in theoretic studies (cf., e.g., \cite{Penrose2003, Stauffer2012}) and mandates
the existence of an edge between two geometrically positioned vertices if and
only if, in our terms, they are no farther apart from each other than the
distance $R_c$.

For simplicity's sake, in this work all agents are assumed to be identical,
thence the values of $R_s$ and $R_c$ are the same for all agents. Additionally,
we note that, as will become clear in the sequel, agents will tend to stand
still at their current locations for far longer than the time they spend moving.
For this reason, another simplifying assumption we make is that, whenever agents
move, they do so instantaneously.

\section{Proposed solution}\label{sec:solution}

At any given time, the probability that an event occurring at that time (or
earlier by no more than $\mathit{VisTime}$ time units) at point $q\in\Omega$ is
detected by at least one of $N$ agents positioned inside $\Omega$ at points
given by the tuple
$s = (s_1,\ldots,s_N) = ((s_{1x},s_{1y}),\dots,(s_{Nx},s_{Ny}))$ is
\begin{equation}
	P(q,s) = 1 - \prod_{k=1}^{N} [1 - p_k(q)].
\end{equation}
Thus, given that an event occurred in $\Omega$ at a certain time $t$ (or
remained visible through time $t$, having occurred no earlier than
$t-\mathit{VisTime}$), the probability that it is detected by at least one agent
is proportional to
\begin{equation}
	\label{eq:eq-4}
	F(s) = \int_{\Omega} \phi(q,t)P(q,s) \mathrm{d}q,
\end{equation}
where the equality holds only if the density $\phi(q,t)$ driving the occurrence
of events at time $t$ can be assumed to vary negligibly with time within a
window of width $\mathit{VisTime}$ inside which instant $t$ is to be found. We
henceforth assume that this is the case.

The importance of function $F$ resides in the following. Should an agent, say
$i$, decide to change its location $s_i$ in order to maximize the probability of
detecting events generated according to the density $\phi$, its best course of
action is to move in the direction given by $\partial F/\partial s_{ix}$ and
$\partial F/\partial s_{iy}$, the components of the gradient of $F$ relative to
agent $i$'s coordinates. We make use of this gradient in what follows, but defer
its calculation to Appendix~\ref{sec:appendixA}.

We now discuss the main elements of our distributed algorithm.

\subsection{Execution modes}

At any point in time, an agent is necessarily engaged in behaving according to
one of two execution modes, either the random mode or the gradient mode. While
in random mode, the agent moves about $\Omega$ somewhat haphazardly, similarly
to what happens in Brownian motion, regardless of the density $\phi$. An agent
in this mode is free to roam the target region and, while doing so, is capable
of putting together an estimate of the event density within a certain vicinity
of each particular point in space and time it goes through. Its contribution to
the overall sensing task is then invaluable: because no agent has any knowledge
of the event density initially, all that can be apprehended about this density
is what agents discover and communicate to one another. And because event
density may change with time, in many situations at least some agents must be
operating in random mode at all times.

If all agents operated in random mode at all times, clearly the entirety of
$\Omega$ would be sensed approximately uniformly, with places of negligible
event density receiving on average as much attention as those of high event
density. The result would be an insufficient concentration of agents in these
latter places, potentially leading them toward being severely undermonitored.
At all times, then, at least some agents must seek to move in the direction of
places of high event density. This is achieved by letting these agents behave in
the gradient mode, which directs agent $i$, when at location $s_i$, to move in
the direction given by $\partial F/\partial s_{ix}$ and
$\partial F/\partial s_{iy}$, i.e., the direction in which the ascent on
$F$ is, locally and at the current time, estimated to be steepest.

\begin{figure}[t]
	\centering
\begin{subfigure}[b]{0.45\textwidth}
	\includegraphics[width=\textwidth]{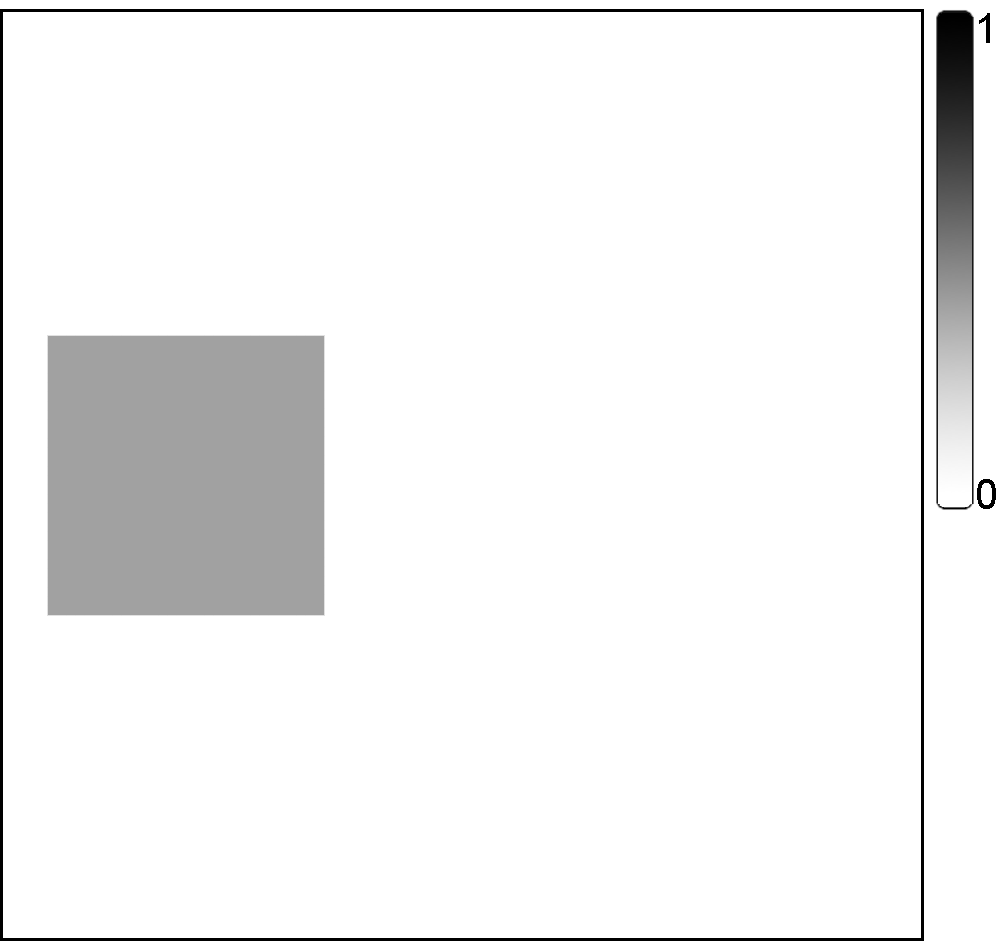}
	\caption{$0\le t\le 60$}
	\label{fig:1a}
\end{subfigure}
\begin{subfigure}[b]{0.45\textwidth}
	\includegraphics[width=\textwidth]{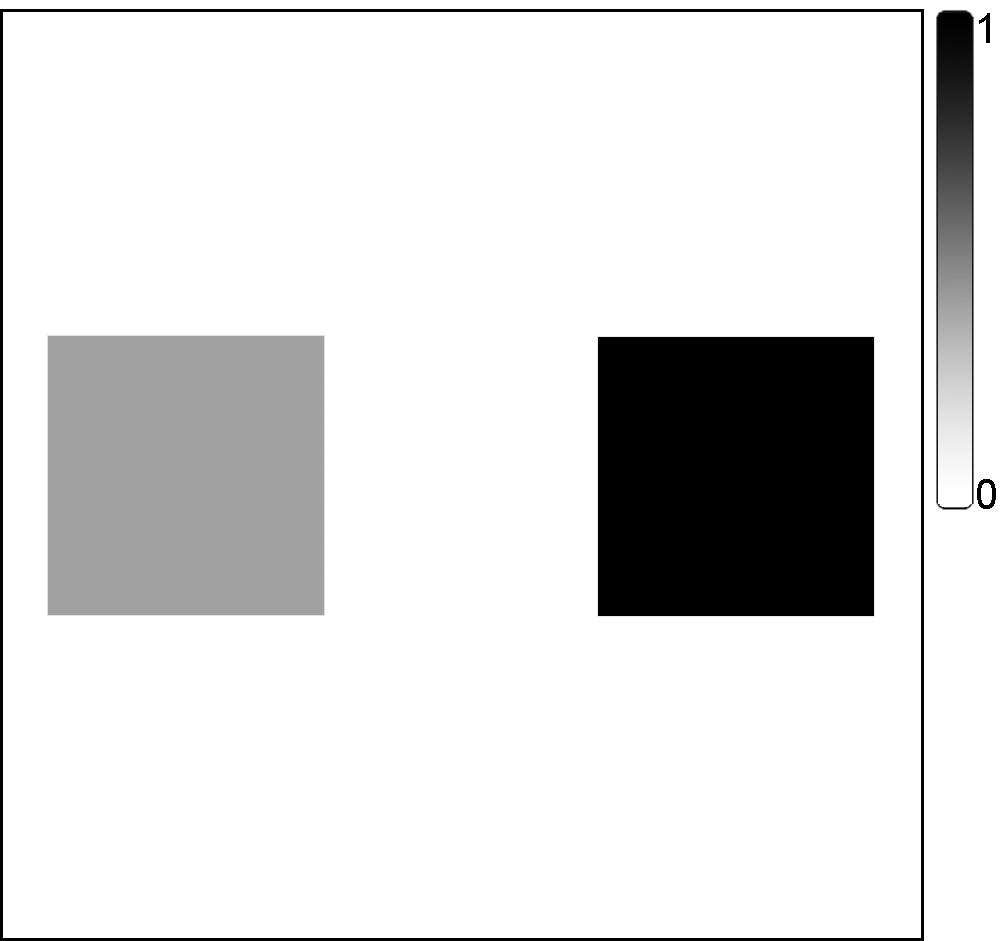}
	\caption{$60<t\le 120$}
	\label{fig:1b}
\end{subfigure}
	\caption{Dynamic event density.}
	\label{fig:1}
\end{figure}

For completeness, we remark that letting all agents operate in gradient mode at
all times is as unreasonable as letting them all be in random mode all the time.
To make the point as clearly as possible, consider the two panels in
Figure~\ref{fig:1}. In each panel the outermost square represents the target
region $\Omega$ while the innermost squares stand for regions with nonzero
event density inside $\Omega$, as indicated by the gray scale. Each panel refers
to a specific time window, $t\in[0,60]$ in panel~(a), $t\in(60,120]$ in
panel~(b). If during the former interval all agents happened to be concentrated
near the light-gray square, and if the two gray squares were sufficiently
separated so that density information could never be relayed across such a
distance, then the higher event densities appearing only during the latter
interval would go permanently unnoticed by lack of at least some agents being
able to move around, possibly toward those densities, while in random mode.

Both the random mode and the gradient mode depend on a $\mathit{StepSize}$
parameter. Each step taken by an agent in random mode carries it
$\mathit{StepSize}$ distance units in a randomly chosen direction. If in
gradient mode, the distance traversed by the agent in the
$(\partial F/\partial s_{ix},\partial F/\partial s_{iy})$ direction in one step
is
$\mathit{StepSize}\vert(\partial F/\partial s_{ix},\partial F/\partial s_{iy})\vert$,
where $i$ is the agent in question. Note that, unlike what happens in random
mode, the distance traversed in one step by agent $i$ in gradient mode can vary
with both $s$ and $t$, as it depends on the magnitude of
$(\partial F/\partial s_{ix},\partial F/\partial s_{iy})$.

\subsection{Saving energy}

Whatever energy an agent has left at any time must be used for moving,
communicating, and sensing. Even though we do not model energy consumption
explicitly, we have striven for our algorithm to be mindful of the agents'
energy usage so that each agent can keep on monitoring its surroundings for
events of interest for as long as possible. We let agents do their sensing at
all times, but moving and communicating are constrained for the sake of saving
energy.

The constraint we impose on an agent's movements is to allow it to move only
once every $\mathit{StillTime}$ time units, regardless of which execution
mode it happens to be operating in, where $\mathit{StillTime}$ is a parameter.
As mentioned earlier, we assume that $\mathit{StillTime}$ is much greater than
the time an agent would spend in a single movement, so agent displacements are
assumed to be instantaneous. As for constraining an agent's ability to
communicate to save energy, we assume that whatever information the agent has
gathered since last moving is accumulated and sent out in a single message right
after it moves next.

\subsection{Message handling}

Each message contains its sender's unique identification and current location,
as well as all events sensed by the sending agent during the past
$\mathit{StillTime}$ time units. Each of these events is represented by the
point in $\Omega$ and the time at which it occurred. Upon receiving any such
message, and having made sure that originally it was sent out by another agent
and is now being received for the first time, an agent passes it on and updates
its current view of the system. This view includes node locations and event
information, the latter easily kept free of duplicates.

Our derivation in Appendix~\ref{sec:appendixA} leaves it clear that, in order
for agent $i$ to compute
$\partial F / \partial s_{ix}$ and $\partial F / \partial s_{iy}$ when operating
in gradient mode, it must make use of all constituents of its current view,
including the locations of agents operating in random mode. However, we know
from experience that such locations can be somewhat jittery and thereby
contribute to gradient calculation mainly as noise. For this reason, agents
operating in random mode always communicate invalid coordinates when sending out
their locations. They can then be easily identified by the receiving agents and
their locations ignored when computing the gradient of $F$.

\subsection{Estimating the event density function}\label{sec:function}

Computing $\partial F / \partial s_{ix}$ and $\partial F / \partial s_{iy}$ also
requires agent $i$ to estimate the value of $\phi$ all over $\Omega$ based on a
recent time window. At time $t$ this window is given by the interval
$[t-\mathit{TimeWindow},t]$, where $\mathit{TimeWindow}$ is another parameter,
one that gives rise to a trade-off between how many events can be used to
estimate $\phi$ (larger values for $\mathit{TimeWindow}$ are preferable) and how
up-to-date $\phi$ can be taken to be (smaller values for $\mathit{TimeWindow}$
are preferable).

In order to handle the spatial dependencies of $\phi$, we subdivide the target
region into $\Delta\times\Delta$ cells for suitably chosen $\Delta$ and use a
new density function, $\phi'$, as a surrogate for $\phi$. The new function is
best written as $\phi'_i(\ell,m,t)$, where $\ell$ and $m$ are nonnegative
integers serving to index each cell along the two dimensions and the subscript
$i$ is meant to emphasize that $\phi'_i$ refers to agent $i$'s local estimate of
$\phi'$. The value of $\phi'_i(\ell,m,t)$ is computed by first totaling the
number of events that according to agent $i$'s current view occurred at some
point in cell $(\ell,m)$ during the time interval $[t-\mathit{TimeWindow},t]$,
and then normalizing the result over all cells so that the greatest resulting
value of $\phi'_i$ is $1$.

\subsection{Switching the execution mode}

At time $t=0$ it is impossible to calculate $\phi'$ for any cell, so all agents
start out in random mode. From then on all agents may switch back and forth
between the two execution modes, depending on the event-occurrence pattern they
detect at their current locations. This movement between the two modes depends
on the current value of 
$\vert(\partial F/\partial s_{ix},\partial F/\partial s_{iy})\vert$, in the case
of agent $i$, and on two further parameters, $\mathit{RtoGMinGrad}$ and
$\mathit{GtoRMaxGrad}$.

Whenever agent $i$ is up for another move, it computes the magnitude
$\vert(\partial F/\break\partial s_{ix},\partial F/\partial s_{iy})\vert$
%$\vert(\partial F/\partial s_{ix},\partial F/\partial s_{iy})\vert$
based on its
current view of the system. If agent $i$ is currently operating in random mode
and the computed magnitude falls strictly above $\mathit{RtoGMinGrad}$, then it
switches to gradient mode already for the move it is about to make. If its
current mode is the gradient mode and the magnitude falls strictly below
$\mathit{GtoRMaxGrad}$, then it switches to random mode.

This is the basic switch policy that agent $i$ follows, but note that it may
perpetuate the agent's operation in gradient mode should it continually
encounter sufficiently high event densities. This can be problematic and a case
in point is precisely the scenario illustrated above with respect to
Figure~\ref{fig:1}. We help agents avoid such a pitfall by employing yet another
parameter, $\mathit{GtoRProb}$, which is the probability with which an agent in
gradient mode switches to random mode upon preparing to make a new move even if
the relevant gradient components indicate its permanence in gradient mode. Upon
switching to random mode, the first $\mathit{GtoRFirstSteps}$ steps are all in
the same, randomly chosen direction, $\mathit{GtoRFirstSteps}$ being another
parameter. The goal of these first steps is to jump-start the agent on the task
of seeking new places to monitor.

\section{Computational experiments}\label{sec:experiments}

In this section we present computational results for a few representative
sensing tasks. They were designed to highlight the working of our algorithm's
innermost core, which is the agents' ability to switch between two execution
modes depending on parameter values. Our results are derived from computational
simulations carried out for $\Omega$ a $1\,000\times1\,000$ square and
$\Delta=10$, with radii $R_s=100$ and $R_c=200$. For each specific application,
supplementary videos are also provided.

We evaluate our algorithm's performance through the use of two metrics, both
related to the number of events that, throughout the entire simulation, get
detected by the agents. Both metrics are significant, but each one's particular
importance in practice depends on the sensing task at hand. They are:
\begin{description}
\item{\textbf{Global fraction of events:}} The fraction representing how many
events were detected by at least one agent.
\item{\textbf{Average local fraction of events:}} The average of several
fractions, one for each agent, each representing how many events were taken
notice of by that specific agent. An agent takes notice of an event by either
detecting it directly or receiving it in a message from another agent.
\end{description}

The first of these metrics is of a global character and tends to be higher as
more events fall within an agent's sensing radius. The second metric is of an
entirely local nature, since it relates to how the various agents perceive the
occurrence of events. This metric tends to be higher as both sensing and
communicating are carried out more effectively by the agents. In what follows,
these two metrics are always reported as averages (with confidence intervals at
the $95\%$ level) over all instants of $200$ independent simulations, each one
starting with the agents positioned uniformly at random inside $\Omega$.

Window-based versions of these metrics can be obtained by taking into account
only the events detected within a certain time window centered at time $t$. We
give plots of these versions in Appendix~\ref{sec:appendixB}, where averages
over the $200$ simulations are shown as a function of $t$, again with confidence
intervals at the $95\%$ level. These intervals are sometimes quite small,
therefore imperceptible in the figures.

In all cases we report results for three distinct agent behaviors regarding the
switch between the random mode and the gradient mode, each one controlled by the
$\mathit{RtoGMinGrad}$, $\mathit{GtoRMaxGrad}$, and $\mathit{GtoRProb}$
parameters:
\begin{description}
\item{\textbf{Random behavior:}} This agent behavior occurs when
$\mathit{RtoGMinGrad}$ is very large (denoted by $\mathit{RtoGMinGrad}=\infty$)
and therefore no agent ever leaves the random mode. This is our control scenario
for evaluating how effective gradient-following can be.
\item{\textbf{Gradient behavior:}} At the opposing end we have
$\mathit{GtoRProb}=0$, in which case an agent can only return to the random mode
as a function of the gradient of $F$. Because all agents start out in random
mode, this means that any agent encountering sufficient event density for it to
switch to gradient mode (as controlled by the $\mathit{RtoGMinGrad}$ parameter)
will remain in gradient mode until that density once again becomes insufficient
(as controlled by the $\mathit{GtoRMaxGrad}$ parameter).
\item{\textbf{Mixed behavior:}} This is the general behavior described in
Section~\ref{sec:solution}. It corresponds to setting $\mathit{RtoGMinGrad}$,
$\mathit{GtoRMaxGrad}$, and $\mathit{GtoRProb}$ to nonzero finite values.
\end{description}

\subsection{Experiment~1: One moving rain cloud}

\begin{figure}[p]
	\centering
\begin{subfigure}[b]{0.45\textwidth}
	\includegraphics[width=\textwidth]{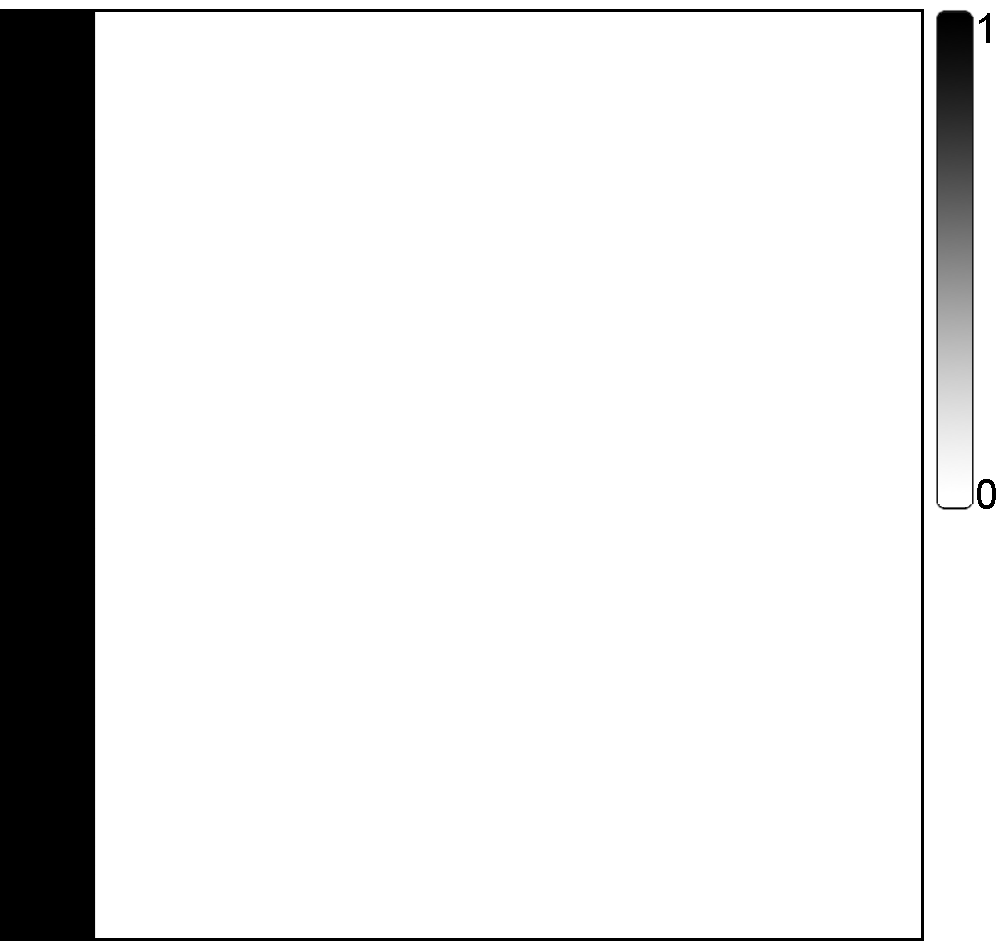}
	\caption{$t=0$}
	\label{fig:2a}
\end{subfigure}
\begin{subfigure}[b]{0.45\textwidth}
	\includegraphics[width=\textwidth]{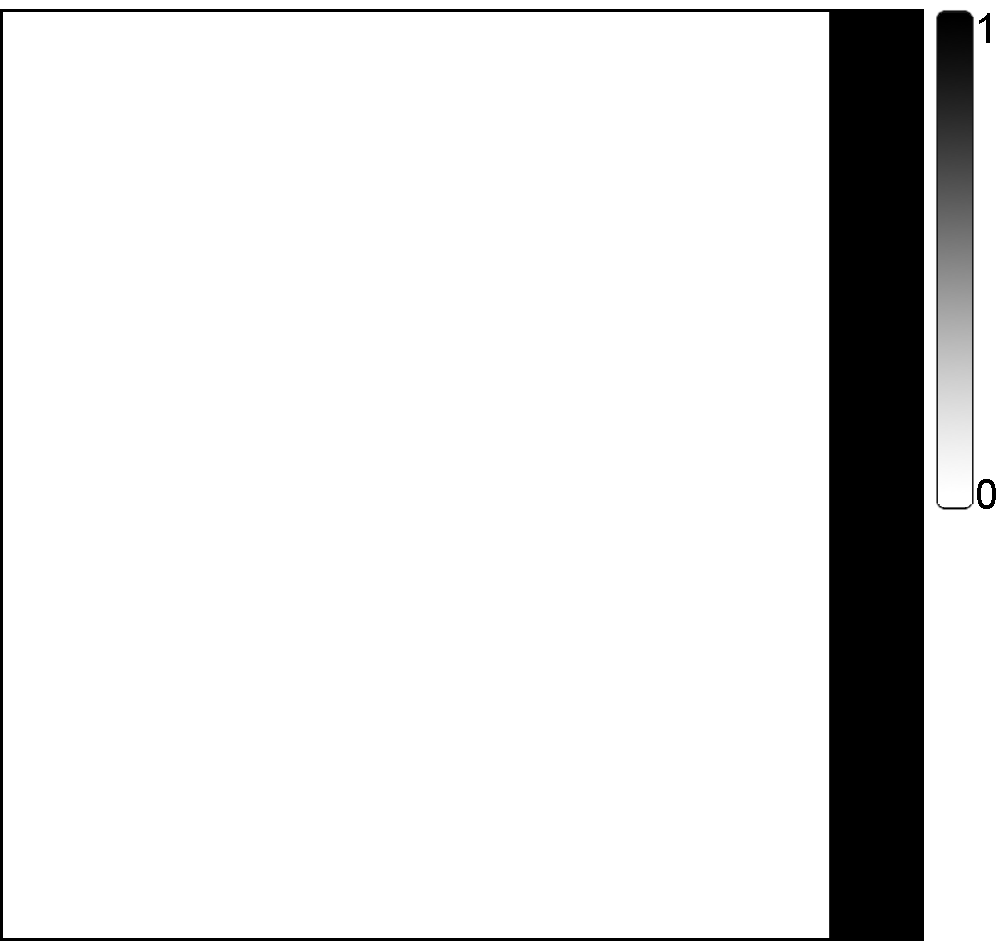}
	\caption{$t=90\,000$}
	\label{fig:2b}
\end{subfigure}
	\caption{Initial (a) and final (b) event densities for Experiment~1.}
	\label{fig:2}
\end{figure}

In this experiment, a rain cloud is simulated that moves uniformly from left to
right across $\Omega$ as shown in Figure~\ref{fig:2}. Each of the simulations
comprised a total of $900\,000$ events, each one representing the fall of a
raindrop on the ground but with $\mathit{VisTime}=0$ (no further visibility, as
in the case of a highly porous surface). All parameters are as given in
Table~\ref{table:1}, results in Table~\ref{table:2}. Sample animations are given
in supplementary videos for the
random (\url{http://www.cos.ufrj.br/\~valmir/EPB2015-E1-R.mp4}),
mixed (\url{http://www.cos.ufrj.br/\~valmir/EPB2015-E1-M.mp4}),
and gradient (\url{http://www.cos.ufrj.br/\~valmir/EPB2015-E1-G.mp4})
behaviors.

\begin{table}[p]
	\centering
	\caption{Parameter values for Experiment~1.}
	\label{table:1}
	\begin{tabular}{|l|c|c|c|}
	\hline
	\multirow{2}{*}{Parameter} 	& \multicolumn{3}{c|}{Behavior}                                                           		\\ \cline{2-4} 
								& \multicolumn{1}{c|}{Random}	& \multicolumn{1}{c|}{Mixed}	& \multicolumn{1}{c|}{Gradient} \\ \hline
	Maximum value of $t$		& \multicolumn{3}{c|}{$90\,000$}									                            \\ \hline
	$N$           				& \multicolumn{3}{c|}{30}           							                            	\\ \hline
	$\mathit{StillTime}$		& \multicolumn{3}{c|}{10}			 							                             	\\ \hline
	$\mathit{RtoGMinGrad}$      & $\infty$                    	& 0.01           				& 0.01       	      			\\ \hline
	$\mathit{GtoRMaxGrad}$     	& $\infty$                    	& 0.00001        				& 0.00001	          			\\ \hline
	$\mathit{GtoRProb}$        	& 1                           	& 0.005          				& 0  	              			\\ \hline
	$\mathit{GtoRFirstSteps}$  	& 0                           	& 10             				& 0	            	  			\\ \hline
	$\mathit{StepSize}$         & \multicolumn{3}{c|}{30}           							                              	\\ \hline
	$\mathit{VisTime}$			& \multicolumn{3}{c|}{0}      	     							                              	\\ \hline
	$\mathit{TimeWindow}$		& \multicolumn{3}{c|}{$1\,000$}        							                              	\\ \hline
\end{tabular}
\end{table}

\begin{table}[p]
	\centering
	\caption{Results for Experiment~1. Confidence intervals correspond to
the $95\%$ level.}
	\label{table:2}
	\begin{tabular}{|l|c|c|c|}
	\hline
	\multirow{2}{*}{Metric} 	& \multicolumn{3}{c|}{Behavior}                                                           		\\ \cline{2-4} 
										& Random 	& Mixed 	& Gradient \\ \hline
	Global fraction of events (\%)           & $15.0\pm 0.1$    & $51.5\pm 0.3$    & $76.3\pm 0.1$   \\ \hline
	Average local fraction of events (\%) 	& $6.8\pm 0.1$     & $33.2\pm 0.4$    & $72.0\pm 0.3$   \\ \hline
\end{tabular}

\end{table}

No matter which of the two metrics we concentrate on, Table~\ref{table:2} is
unequivocal in pointing to the gradient behavior as the one to be chosen. In a
situation in which events leave no footprint and moreover occur only on a moving
patch inside $\Omega$, the only possibility for agents operating in random mode
to come across any event at the time of its occurrence is to chance upon the
moving patch and then switch to gradient mode. This explains the superiority of
the gradient behavior, followed by the mixed behavior and by the random behavior
(this one trailing far behind).

Further examining the figures for the gradient behavior reveals only a small
difference between the global metric and the average local one. This means that,
while clustering about the moving patch, those agents that do so are capable of
efficiently communicating to one another most of what they sense. In the end, on
average an agent has been notified of almost all events that got detected by at
least one agent.

Additional details can be seen in Figure~\ref{fig:B1}, where the evolution in
time of the window-based versions of the two metrics is shown for this
experiment. The metrics' evolution is mostly uneventful, with values rising
toward a certain level and being sustained there throughout the simulation. The
one marked exception is that of the window-based local metric when the agents
operate under the mixed behavior. In this case, the value that is attained after
the initial growth fails to be sustained and begins to undergo a slow decrease.
Readily, the farther the rainy patch is from the middle longitudes of $\Omega$,
the less effective inter-agent communication becomes. This is so because the
distance between the agents actually detecting the events (those operating in
gradient mode) and the ones roaming about all of $\Omega$ (those operating in
random mode) becomes on average significantly larger.

\subsection{Experiment~2: Two moving rain clouds}

\begin{figure}[p]
	\centering
\begin{subfigure}[b]{0.45\textwidth}
	\includegraphics[width=\textwidth]{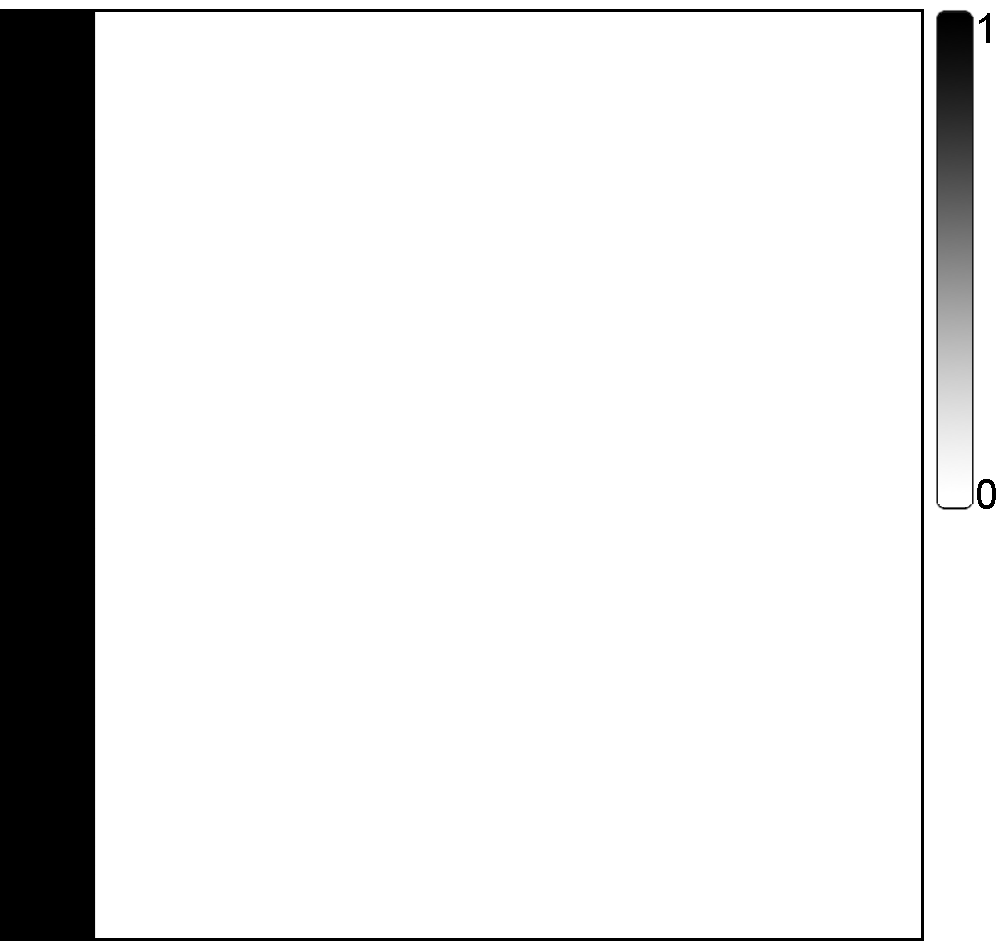}
	\caption{$t=0$}
	\label{fig:3a}
\end{subfigure}
\begin{subfigure}[b]{0.45\textwidth}
	\includegraphics[width=\textwidth]{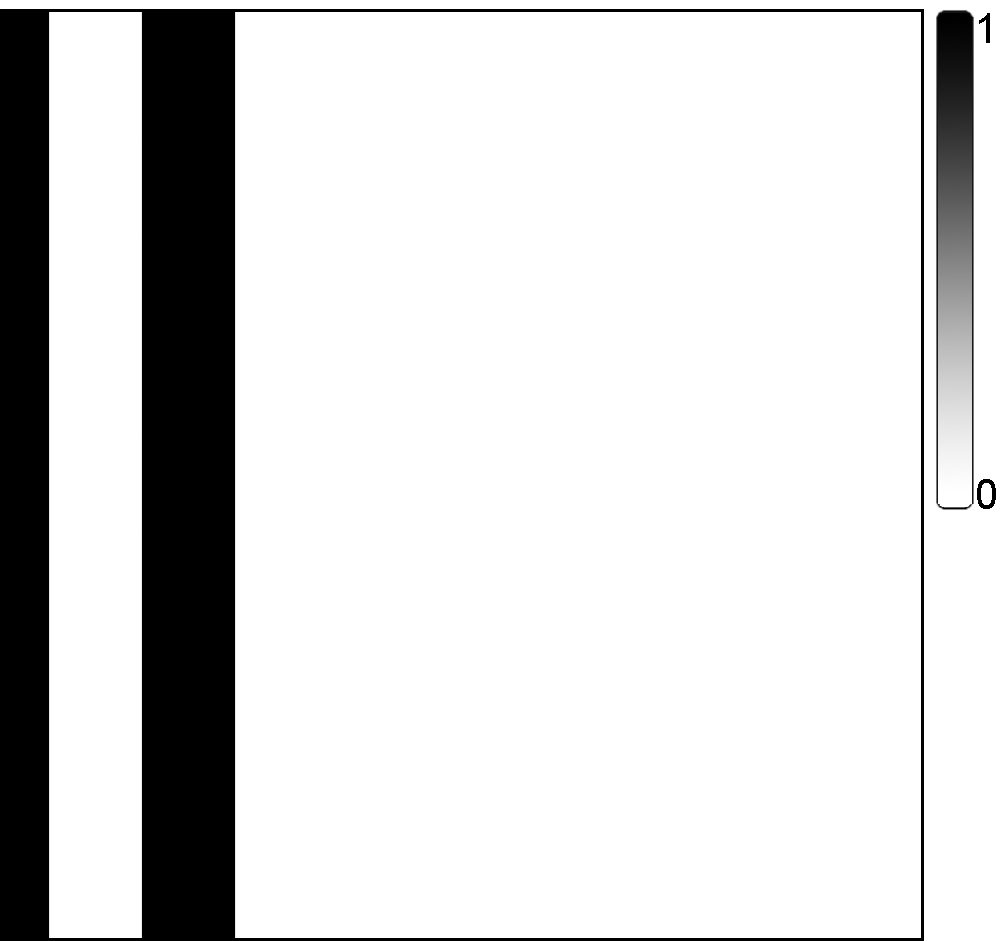}
	\caption{$t=15\,000$}
	\label{fig:3b}
\end{subfigure}
\linebreak\linebreak
\begin{subfigure}[b]{0.45\textwidth}
	\includegraphics[width=\textwidth]{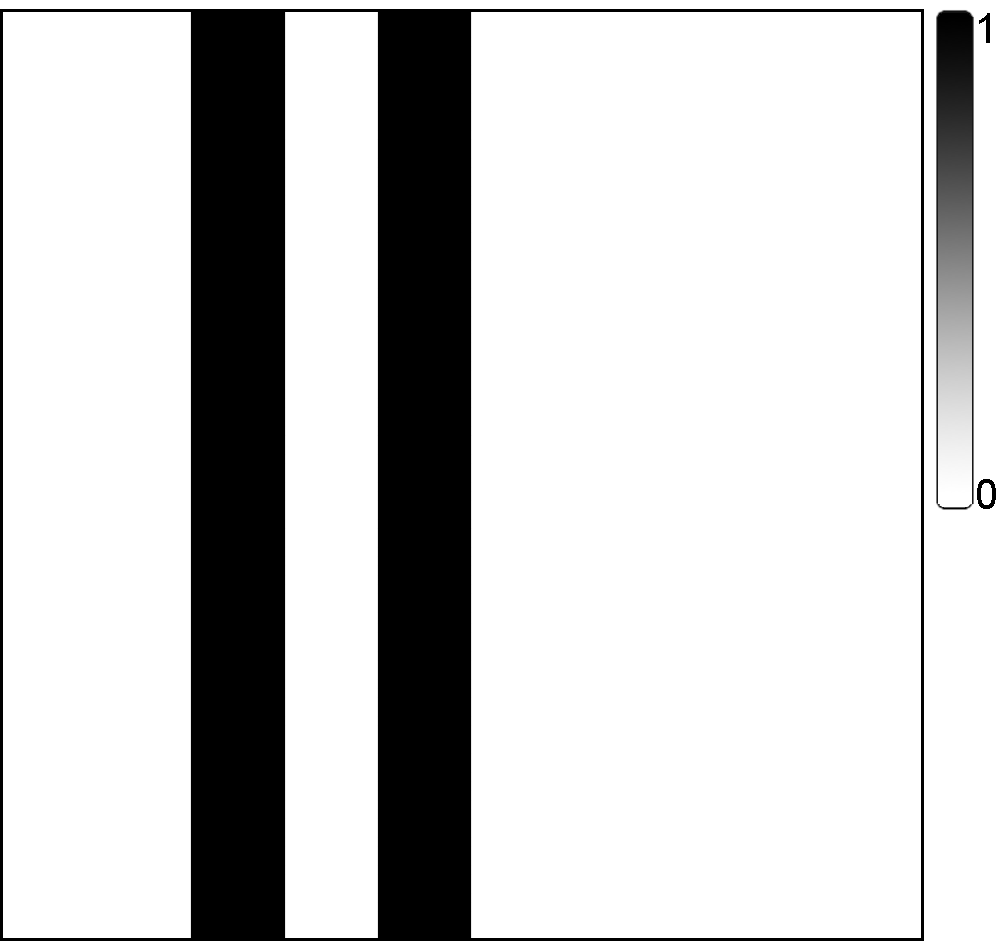}
	\caption{$t=40\,000$}
	\label{fig:3c}
\end{subfigure}
\begin{subfigure}[b]{0.45\textwidth}
	\includegraphics[width=\textwidth]{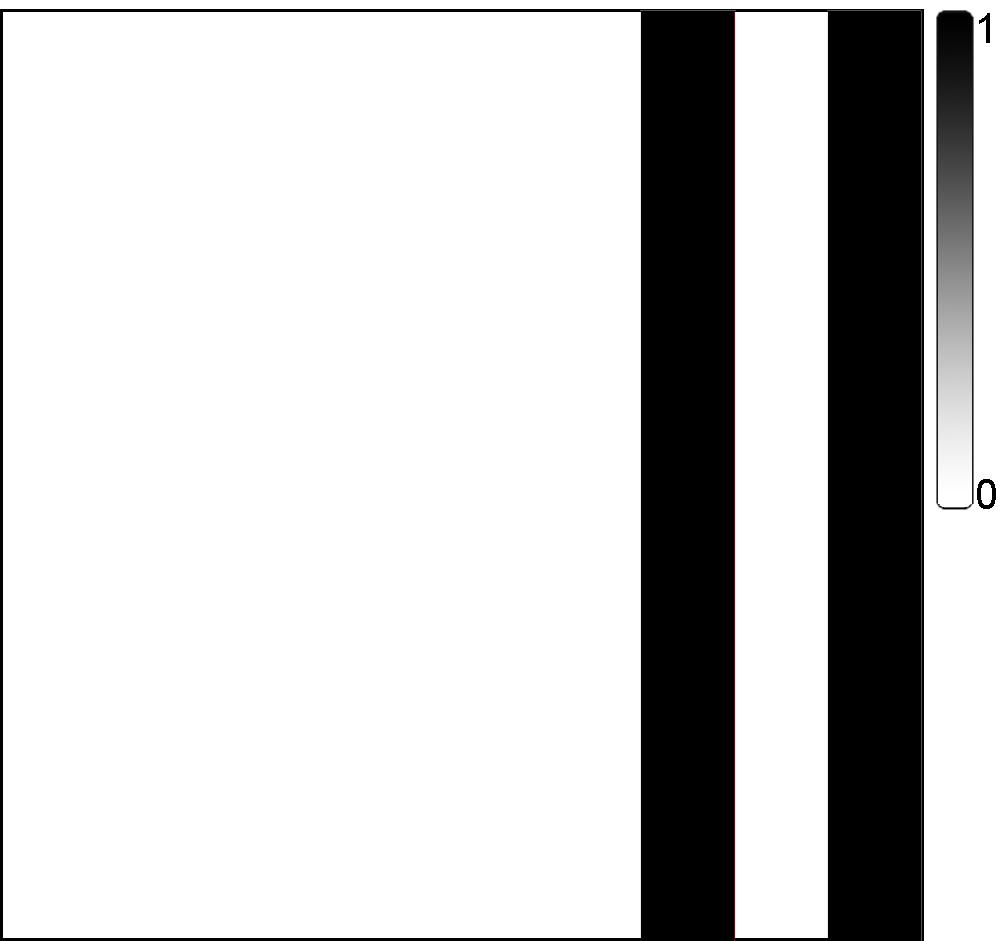}
	\caption{$t=90\,000$}
	\label{fig:3d}
\end{subfigure}
	\caption{Evolution of the event density for Experiment~2 (a--d).}
	\label{fig:3}
\end{figure}

The setting for this second experiment is similar to that of Experiment~1,
the main difference being a second rain cloud that trails the first across
$\Omega$, having entered the scene at $t=10\,000$ (Figure~\ref{fig:3}). The
number of events per simulation now totals $1\,650\,000$. Parameter values are
given in Table~\ref{table:3}, results in Table~\ref{table:4}. Sample animations
are given in supplementary videos for the
random (\url{http://www.cos.ufrj.br/\~valmir/EPB2015-E2-R.mp4}),
mixed (\url{http://www.cos.ufrj.br/\~valmir/EPB2015-E2-M.mp4}),
and gradient (\url{http://www.cos.ufrj.br/\~valmir/EPB2015-E2-G.mp4})
behaviors.

Comparing the results in Table~\ref{table:4} to those of Table~\ref{table:2}
(for Experiment~1) reveals that now agents are better off adopting the mixed
(rather than the gradient) behavior. To see why this is so, consider that the
appearance of the second rain cloud is challenging to the agents currently
concentrating on the first one, in the sense that, in order to reach it, they
first have to leave the gradient mode, then cross the dry patch between the two
rain clouds, and finally revert to gradient mode once again upon detecting the
increase in event density that the second rain cloud entails.

The superiority of the mixed behavior in the case of Experiment~2 holds for the
two metrics we have adopted, with the gradient behavior following close behind
and the random behavior once again trailing significantly farther back.
Moreover, once again it is the case that, for the mixed behavior, the two
metrics do not differ greatly and therefore we see that whatever the agents
detect is nearly as effective locally as it is globally. In the case of
Experiment~2, this is true of the gradient behavior as well.

\begin{table}[t]
	\centering
	\caption{Parameters values for Experiment~2.}
	\label{table:3}
	\begin{tabular}{|l|c|c|c|}
	\hline
	\multirow{2}{*}{Parameter} 	& \multicolumn{3}{c|}{Behavior}                                                           		\\ \cline{2-4} 
								& \multicolumn{1}{c|}{Random}	& \multicolumn{1}{c|}{Mixed}	& \multicolumn{1}{c|}{Gradient} \\ \hline
	Maximum value of $t$		& \multicolumn{3}{c|}{$90\,000$} 									                            \\ \hline
	$N$           				& \multicolumn{3}{c|}{30}           							                              	\\ \hline
	$\mathit{StillTime}$		& \multicolumn{3}{c|}{10}			 							                              	\\ \hline
	$\mathit{RtoGMinGrad}$      & $\infty$                      & 0.01                          & 0.01       	      			\\ \hline
	$\mathit{GtoRMaxGrad}$     	& $\infty$                      & 0.00001                       & 0.00001	          			\\ \hline
	$\mathit{GtoRProb}$        	& 1                             & 0.0005                        & 0  	              			\\ \hline
	$\mathit{GtoRFirstSteps}$  	& 0                             & 10                            & 0	            	  			\\ \hline
    $\mathit{StepSize}$         & \multicolumn{3}{c|}{30}           							                              	\\ \hline
	$\mathit{VisTime}$			& \multicolumn{3}{c|}{0}             							                              	\\ \hline
	$\mathit{TimeWindow}$		& \multicolumn{3}{c|}{$1\,000$}          							                            \\ \hline
\end{tabular}
\end{table}

\begin{table}[t]
	\centering
	\caption{Results for Experiment~2. Confidence intervals correspond to
the $95\%$ level.}
	\label{table:4}
	\begin{tabular}{|l|c|c|c|}
	\hline
	\multirow{2}{*}{Metric} 	& \multicolumn{3}{c|}{Behavior}                                                           		\\ \cline{2-4} 
										& Random 	& Mixed 	& Gradient \\ \hline
	Global fraction of events (\%)           & $14.9\pm 0.1$    & $50.1\pm 0.2$    & $41.5\pm 0.1$   \\ \hline
	Average local fraction of events (\%) 	& $6.9\pm 0.1$     & $44.8\pm 0.3$    & $39.1\pm 0.1$   \\ \hline
\end{tabular}

\end{table}

Still in comparison to the data for Experiment~1, it is worth noting that
figures are now significantly lower under the gradient behavior (this, in fact,
is why in Experiment~2 the mixed behavior rose to the top). The reason for this,
clearly, is that while in the gradient behavior agents neglect the second rain
cloud altogether, thence the ensuing drop in their performance.

Additional insight can be gained by examining the evolution in time of the
window-based versions of the two metrics (Figure~\ref{fig:B2}). Clearly, up
until $t=10\,000$ (the time of appearance of the second rain cloud), the
agents' performances in either of the two gradient-based behaviors are
indistinguishable. Between this time and that at which the second wet patch
becomes fully contained in $\Omega$, performance under these two behaviors
undergoes a significant drop. Afterwards the mixed behavior succeeds in
gradually leading the system to some degree of recovery from the drop while the
gradient behavior does not.

\subsection{Experiment 3: Persistent pollutants}

\begin{figure}[p]
	\centering
\begin{subfigure}[b]{0.45\textwidth}
	\includegraphics[width=\textwidth]{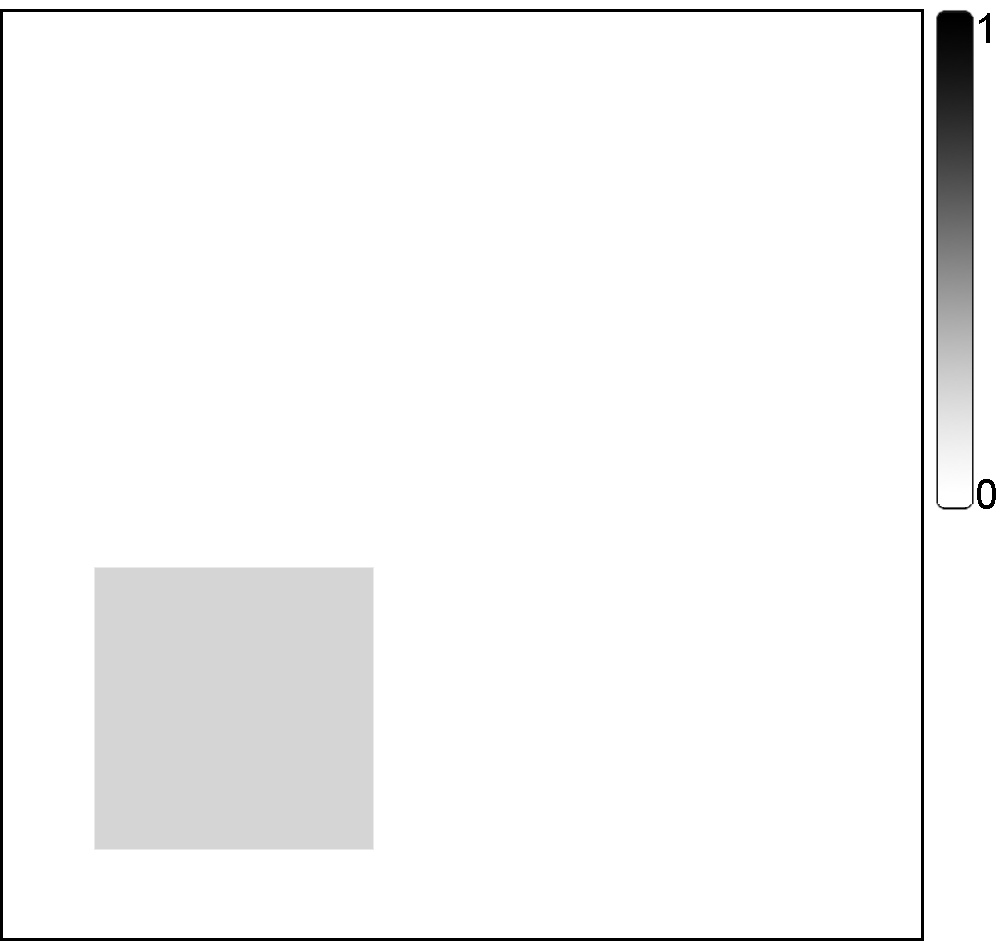}
	\caption{$0\le t\le 25\,000$}
	\label{fig:4a}
\end{subfigure}
\begin{subfigure}[b]{0.45\textwidth}
	\includegraphics[width=\textwidth]{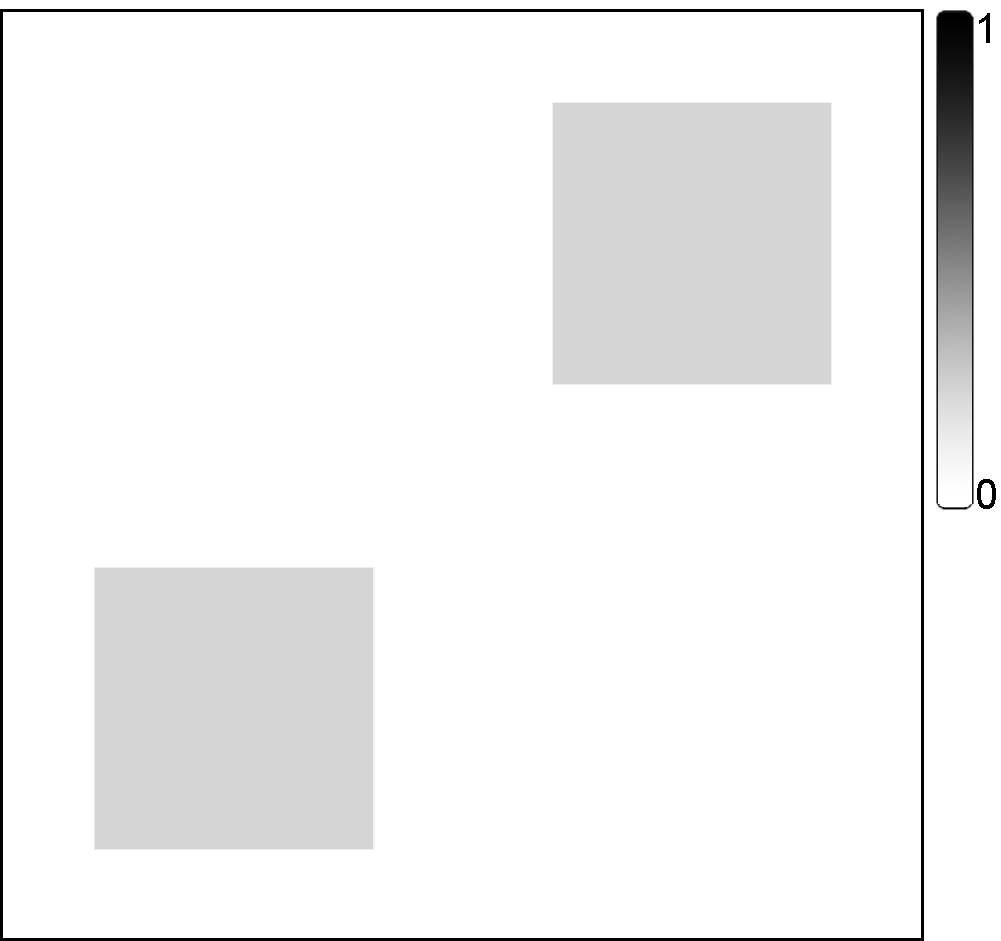}
	\caption{$25\,000<t\le 50\,000$}
	\label{fig:4b}
\end{subfigure}
\linebreak\linebreak
\begin{subfigure}[b]{0.45\textwidth}
	\includegraphics[width=\textwidth]{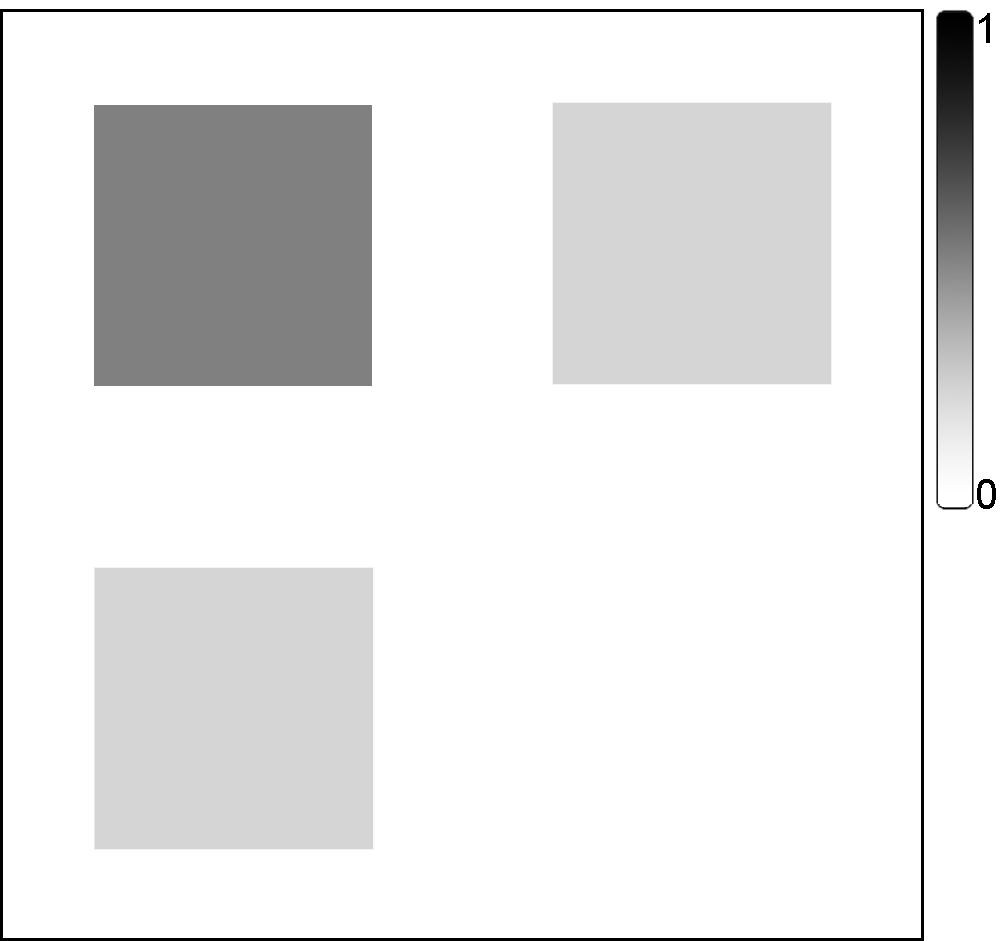}
	\caption{$50\,000<t\le 75\,000$}
	\label{fig:4c}
\end{subfigure}
\begin{subfigure}[b]{0.45\textwidth}
	\includegraphics[width=\textwidth]{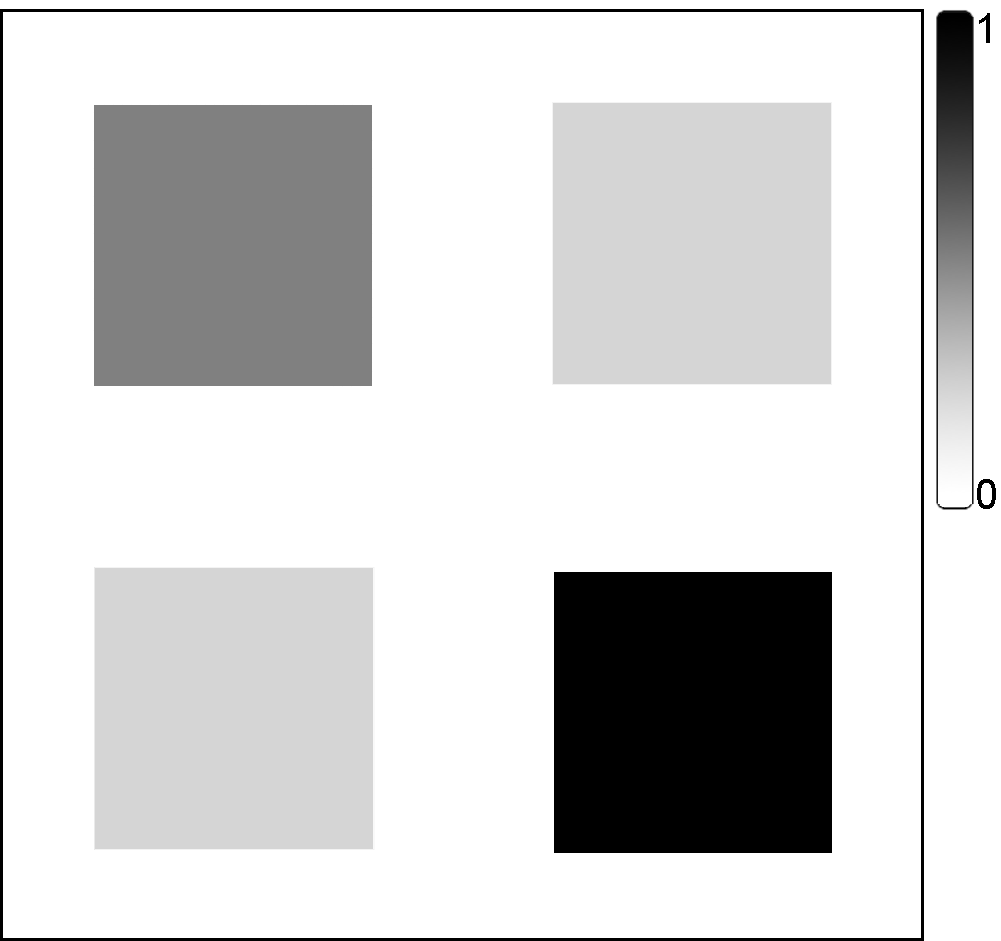}
	\caption{$75\,000<t\le 100\,000$}
	\label{fig:4d}
\end{subfigure}
	\caption{Evolution of the event density for Experiment~3 (a--d).}
	\label{fig:4}
\end{figure}

\begin{table}[t]
	\centering
	\caption{Parameter values for Experiment~3.}
	\label{table:5}
	\begin{tabular}{|l|c|c|c|}
	\hline
	\multirow{2}{*}{Parameter} 	& \multicolumn{3}{c|}{Behavior}                                                           		\\ \cline{2-4} 
								& \multicolumn{1}{c|}{Random}	& \multicolumn{1}{c|}{Mixed}	& \multicolumn{1}{c|}{Gradient} \\ \hline
	Maximum value of $t$		& \multicolumn{3}{c|}{$100\,000$} 									                            \\ \hline
	$N$           				& \multicolumn{3}{c|}{50}           							                              	\\ \hline
	$\mathit{StillTime}$		& \multicolumn{3}{c|}{20}			 							                              	\\ \hline
	$\mathit{RtoGMinGrad}$      & $\infty$                      & 0.01                          & 0.01 	              			\\ \hline
	$\mathit{GtoRMaxGrad}$     	& $\infty$                      & 0.00001                       & 0.00001	          			\\ \hline
	$\mathit{GtoRProb}$        	& 1                             & 0.01                          & 0          	      			\\ \hline
	$\mathit{GtoRFirstSteps}$  	& 0                             & 10                            & 0              	  			\\ \hline
	$\mathit{StepSize}$         & \multicolumn{3}{c|}{25}           							                              	\\ \hline
	$\mathit{VisTime}$			& \multicolumn{3}{c|}{100}           							                              	\\ \hline
	$\mathit{TimeWindow}$		& \multicolumn{3}{c|}{$1\,000$}          							                            \\ \hline
\end{tabular}
\end{table}

\begin{table}[t]
	\centering
	\caption{Results for Experiment~3. Confidence intervals correspond to
the $95\%$ level.}
	\label{table:6}
	\begin{tabular}{|l|c|c|c|}
	\hline
	\multirow{2}{*}{Metric} 	& \multicolumn{3}{c|}{Behavior}                                                           		\\ \cline{2-4} 
										& Random 	& Mixed 	& Gradient \\ \hline
	Global fraction of events (\%)           & $57.5\pm 0.2$    & $80.6\pm 0.2$    & $45.3\pm 0.3$   \\ \hline
	Average local fraction of events (\%) 	& $51.8\pm 0.3$    & $59.7\pm 0.4$    & $24.5\pm 0.1$   \\ \hline
\end{tabular}

\end{table}

This experiment continues the trend initiated by Experiment~2 in that events
are concentrated in more than one patch inside $\Omega$ and moreover new patches
appear as time elapses. The two settings differ, though, mainly in that the
event patches no longer move, but also in that they now appear more abruptly and
also with different event densities, totaling $750\,000$ events
(Figure~\ref{fig:4}). The new experiment also sets $\mathit{VisTime}=100$, the
metaphor now being that of pollutant particles that stick to the environment and
therefore have a nonzero visibility time. The parameter values used are those
shown in Table~\ref{table:5}. We give results in Table~\ref{table:6}. Sample
animations are given in supplementary videos for the
random (\url{http://www.cos.ufrj.br/\~valmir/EPB2015-E3-R.mp4}),
mixed (\url{http://www.cos.ufrj.br/\~valmir/EPB2015-E3-M.mp4}),
and gradient (\url{http://www.cos.ufrj.br/\~valmir/EPB2015-E3-G.mp4})
behaviors.

The most striking figures in Table~\ref{table:6} are those for the random
behavior, which not only surpass those of the previous experiments by a great
margin but also put random behavior ahead of gradient behavior. The main reason
for this seems clear: given that the event patches remain spatially static once
they appear, the freely roaming agents in random mode eventually succeed in
detecting a sizable fraction of the events, even if they cover the denser
patches somewhat thinly.

Other interesting data in the table are those referring to the agents'
performance under the gradient behavior. In this case they fare worse than they
do under random behavior, and again the reason seems clear: as denser patches
continue to crop up, those agents not yet involved with monitoring the already
existing patches will tend to be too few, certainly fewer than necessary to
provide the new patches with any degree of satisfactory coverage.

Given these explanations for the agents' performance under the random behavior
and the gradient behavior, it should be no surprise that, as in Experiment~2,
once again the mixed behavior is the best option. Its mixed nature seems to
alleviate the shortcomings of the other two behaviors, much as it strengthens
their advantages. Note, additionally, that in the two behaviors involving any
participation of the gradient mode the local metrics are valued significantly
below the global one. What seems to be happening is that, because of the
patches' spatial separation from one another, effectively conveying sensing
information from agents at one patch to those at another depend more on agents
operating in random mode than it would otherwise do.

The evolution in time of the window-based metrics, shown in Figure~\ref{fig:B3},
provides further valuable information. As in the case of Experiment~2, the
appearance of each new, spatially separated patch of events causes significant
drops in performance when the agents are in either of the two gradient-based
behaviors. Recovery always follows under the mixed behavior, though to somewhat
lower levels, which eventually guarantees its overall superiority. As for the
gradient behavior, some recovery also takes place but is severely insufficient
to lead to any significant level of overall performance. In particular, such
recovery only happens while there exist agents operating in random mode, which 
is reflected in the complete absence of any recovery from the appearance of the
new event patches right past $t=50\,000$ and $t=75\,000$.

\section{Conclusion}\label{sec:conclusion}

We have introduced a new distributed algorithm for a collection of mobile
agents to monitor a two-dimensional domain seeking to sense and possibly store
events of a certain nature occurring in their target region. The algorithm is
fully distributed, in the sense that not even leader agents are needed, and is
also fully adaptive, in the sense that it strives for agents to perform well in
their sensing tasks even if the events in question vary both spatially and
temporally. At the heart of the algorithm lies the notion of an execution mode,
of which there are two (the random mode and the gradient mode), which in essence
is a pattern of behavior allowing the agent to concentrate on certain specific
aspects of the monitoring task to be carried out.

Agents operating in random mode are tailored to handle the uncertainties of some
situations, such as those in which events tend to happen unpredictably. Those
operating in gradient mode find guidance in the locally perceived density of
events to move toward points that may surround their current locations having
higher values of a density-dependent function. Depending on a variety of
parameters whose setting in turn depends on the specific application at hand,
agents are capable of switching back and forth between the two execution modes
and thereby help the system as a whole profit as much as possible from each
mode's characteristics.

We illustrated some of this interplay of parameter values by means of three
computational experiments, each one designed to highlight those aspects of an
application that justify the spectrum of behaviors we called random, mixed, and
gradient. Such aspects include the occurrence of events whose spatial
coordinates vary with time, both within a connected portion of the target region
and otherwise; the rapid change in the overall event density across spatially
separated portions of the target region; and some of the peculiarities of an
event's environmental footprint.

Our solution has been given for the case of a two-dimensional target region but
nothing prevents its straightforward extension to three dimensions, provided
only that the agents themselves can exist and move in a three-dimensional
setting. It would also be straightforward to adopt sensing and communication
models more realistic than those of Eqs.~(\ref{eq:eq-1}) and ~(\ref{eq:eq-2}),
respectively. In a similar vein, and notwithstanding its already large set of
parameters, endowing the algorithm with even more parameters should not be out
of the question. In fact, the ability to have so many of its details adjustable
is precisely what makes it potentially applicable to a wide variety of
situations.

\subsection*{Acknowledgments}

The authors acknowledge partial support from CNPq, CAPES, and a FAPERJ BBP
grant.

\bibliography{main}
\bibliographystyle{unsrt}

\appendix
\numberwithin{equation}{section}
\numberwithin{figure}{section}

\newpage
\section{Computing the gradient}\label{sec:appendixA}

We first rewrite Eq.~(\ref{eq:eq-4}) for fixed $t$ as
\begin{equation}
F(s)=
\int_\Omega\phi(q)\mathrm{d}q-
\int_\Omega\phi(q)\prod_{k=1}^N[1-p_k(q)]\mathrm{d}q.
\end{equation}
Then we proceed by differentiating $F$ with respect to $s_{ix}$, which yields
\begin{equation}
\frac{\partial F}{\partial s_{ix}}=
-\int_{\Omega_i}\phi(q)\prod_{{k=1}\atop{k\neq i}}^N[1-p_k(q)]
\frac{\partial}{\partial s_{ix}}[1-p_i(q)]
\mathrm{d}q,
\end{equation}
where $\Omega_i$ is the intersection of $\Omega$ with the radius-$R_s$ circle
centered at $s_i=(s_{ix},s_{iy})$. This follows from the fact that $p_i(q)$
does not depend on $s_{ix}$ for $q\in\Omega\setminus\Omega_i$, by
Eq.~(\ref{eq:eq-1}). By the same token, noting further that $p_k(q)=0$ whenever
$\Omega_k\cap\Omega_i=\emptyset$ leads to
\begin{equation}
\frac{\partial F}{\partial s_{ix}}=
-\int_{\Omega_i}\phi(q)
\prod_{{{k=1}\atop{k\neq i}}\atop{\Omega_k\cap\Omega_i\neq\emptyset}}^N
[1-p_k(q)]
\frac{\partial}{\partial s_{ix}}[1-p_i(q)]
\mathrm{d}q.
\end{equation}
Using $q=(q_x,q_y)$, and recalling that
$d_i=[(s_{ix}-q_x)^2+(s_{iy}-q_y)^2]^{1/2}$, we obtain
\begin{equation}
\frac{\partial p_i}{\partial s_{ix}}=
\frac{2(s_{ix}-q_x)}{R_s}
\left(\frac{1}{R_s}-\frac{1}{d_i}\right)
\end{equation}
and
\begin{equation}
\frac{\partial F}{\partial s_{ix}}=
\int_{\Omega_i}\phi(q)
\prod_{{{k=1}\atop{k\neq i}}\atop{\Omega_k\cap\Omega_i\neq\emptyset}}^N
[1-p_k(q)]
\frac{2(s_{ix}-q_x)}{R_s}
\left(\frac{1}{R_s}-\frac{1}{d_i}\right)
\mathrm{d}q.
\end{equation}

As explained in Section~\ref{sec:function}, for use in the algorithm by agent
$i$ in relation to time $t$, subdividing $\Omega$ into the square cells of side
$\Delta$ while $\phi'_i$ substitutes for $\phi$ requires an approximation to the
above expression for $\partial F/\partial s_{ix}$. Using $q=(q_\ell,q_m)$ as the
central point of cell $(\ell,m)$ leads to the discrete approximation
\begin{equation}
\frac{\partial F}{\partial s_{ix}}\approx
\sum_\ell\sum_m\phi'_i(\ell,m,t)
\prod_{{{k=1}\atop{k\neq i}}\atop{\Omega_k\cap\Omega_i\neq\emptyset}}^N
[1-p_k(q)]
\frac{2(s_{ix}-q_x)}{R_s}
\left(\frac{1}{R_s}-\frac{1}{d_i}\right),
\end{equation}
where $\ell$ and $m$ range over the cells that intersect $\Omega_i$ and
$d_i=[(s_{ix}-q_\ell)^2+(s_{iy}-q_m)^2]^{1/2}$.

The case of $\partial F/\partial s_{iy}$ is entirely analogous and yields
\begin{equation}
\frac{\partial F}{\partial s_{iy}}\approx
\sum_\ell\sum_m\phi'_i(\ell,m,t)
\prod_{{{k=1}\atop{k\neq i}}\atop{\Omega_k\cap\Omega_i\neq\emptyset}}^N
[1-p_k(q)]
\frac{2(s_{iy}-q_y)}{R_s}
\left(\frac{1}{R_s}-\frac{1}{d_i}\right).
\end{equation}

\newpage
\section{Additional figures}\label{sec:appendixB}

The figures in this appendix accompany the discussion in
Section~\ref{sec:experiments}. They refer to the window-based versions of the
two metrics introduced in that section.

\begin{figure}[h]
	\centering
\begin{subfigure}[b]{0.95\textwidth}
	\includegraphics[width=\textwidth]{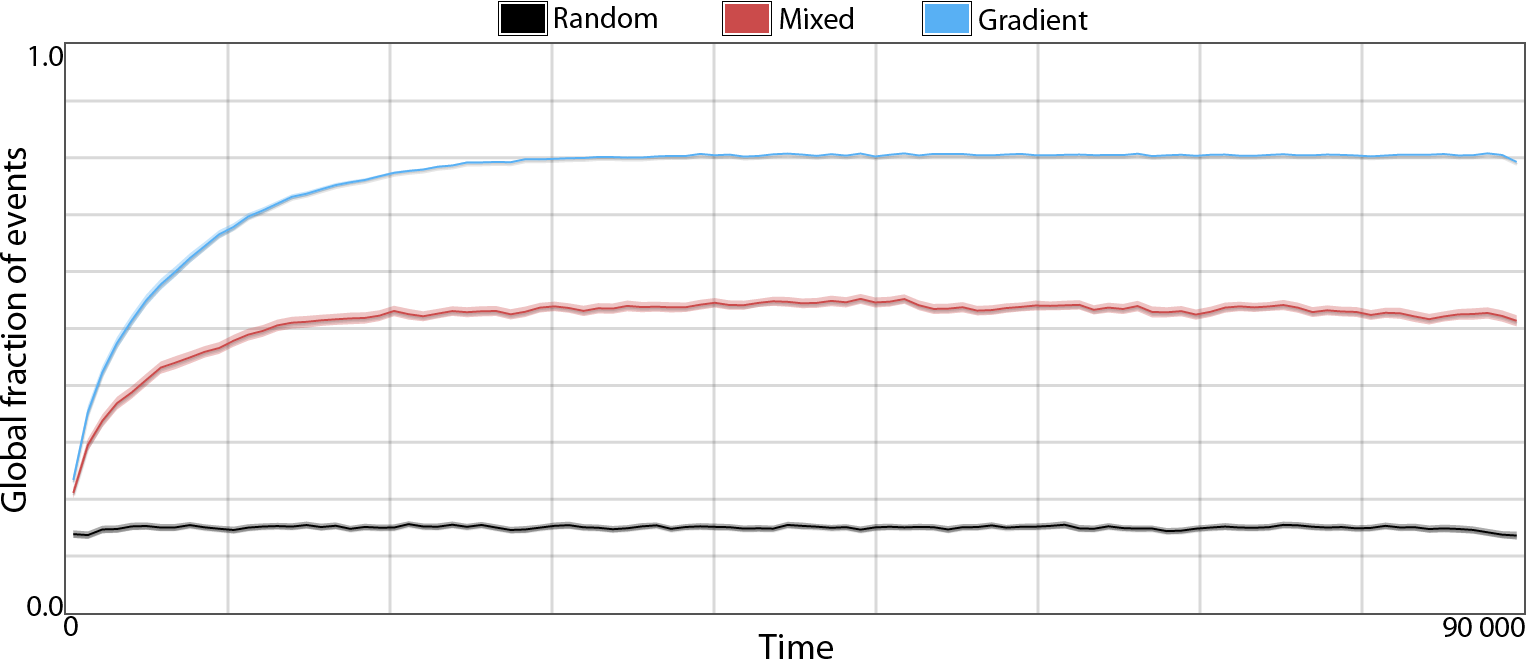}
	\caption{}
	\label{fig:B1a}
\end{subfigure}
\linebreak\linebreak
\begin{subfigure}[b]{0.95\textwidth}
	\includegraphics[width=\textwidth]{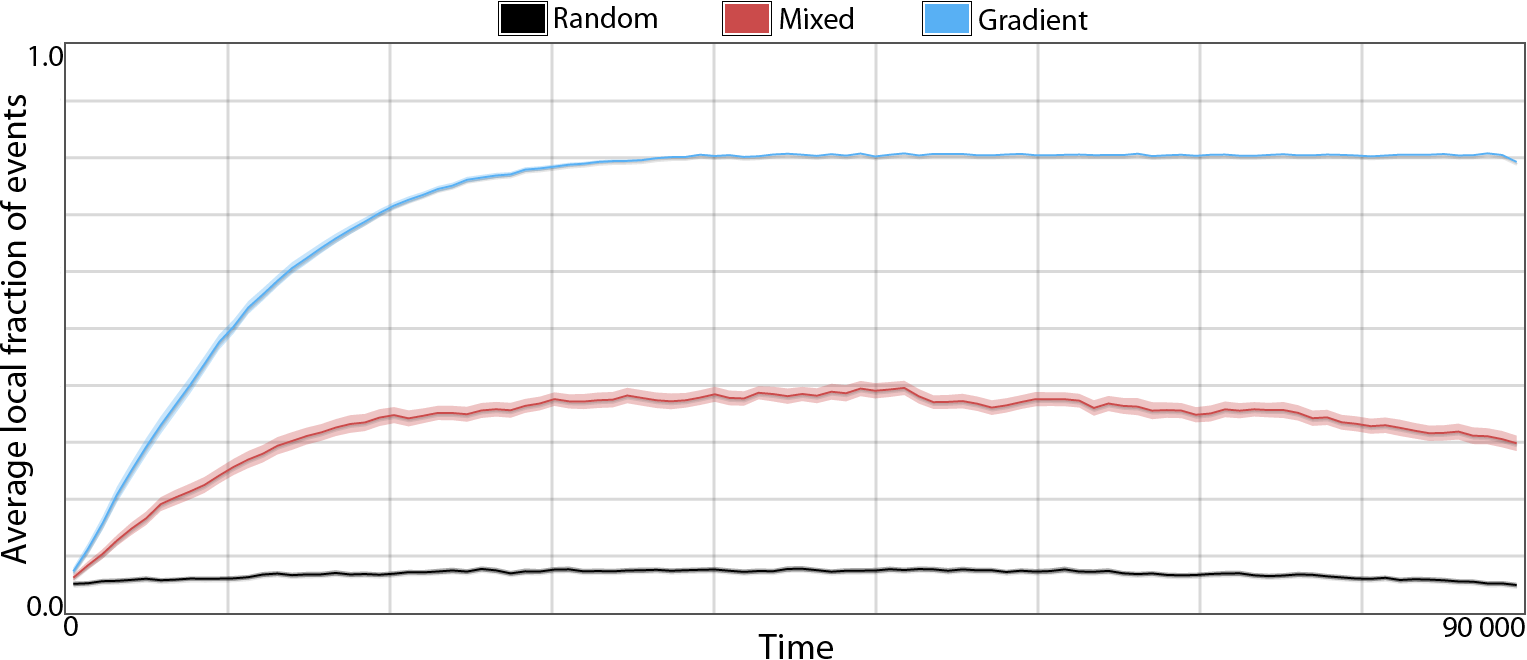}
	\caption{}
	\label{fig:B1b}
\end{subfigure}
	\caption{Evolution in time of the average global (a) and local (b)
window-based metrics for Experiment~1. Window size is $900$. Confidence
intervals correspond to the $95\%$ level.}
	\label{fig:B1}
\end{figure}

\begin{figure}[p]
	\centering
\begin{subfigure}[b]{0.95\textwidth}
	\includegraphics[width=\textwidth]{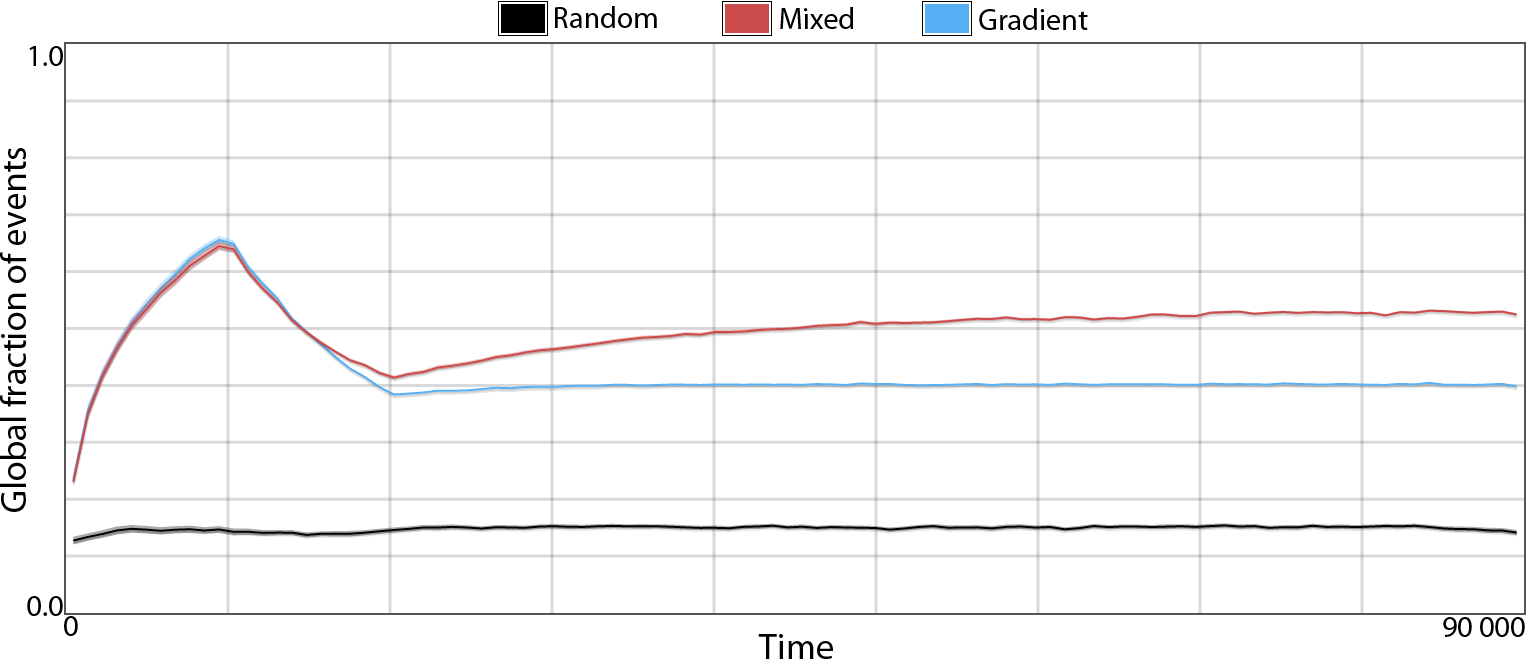}
	\caption{}
	\label{fig:B2a}
\end{subfigure}
\linebreak\linebreak
\begin{subfigure}[b]{0.95\textwidth}
	\includegraphics[width=\textwidth]{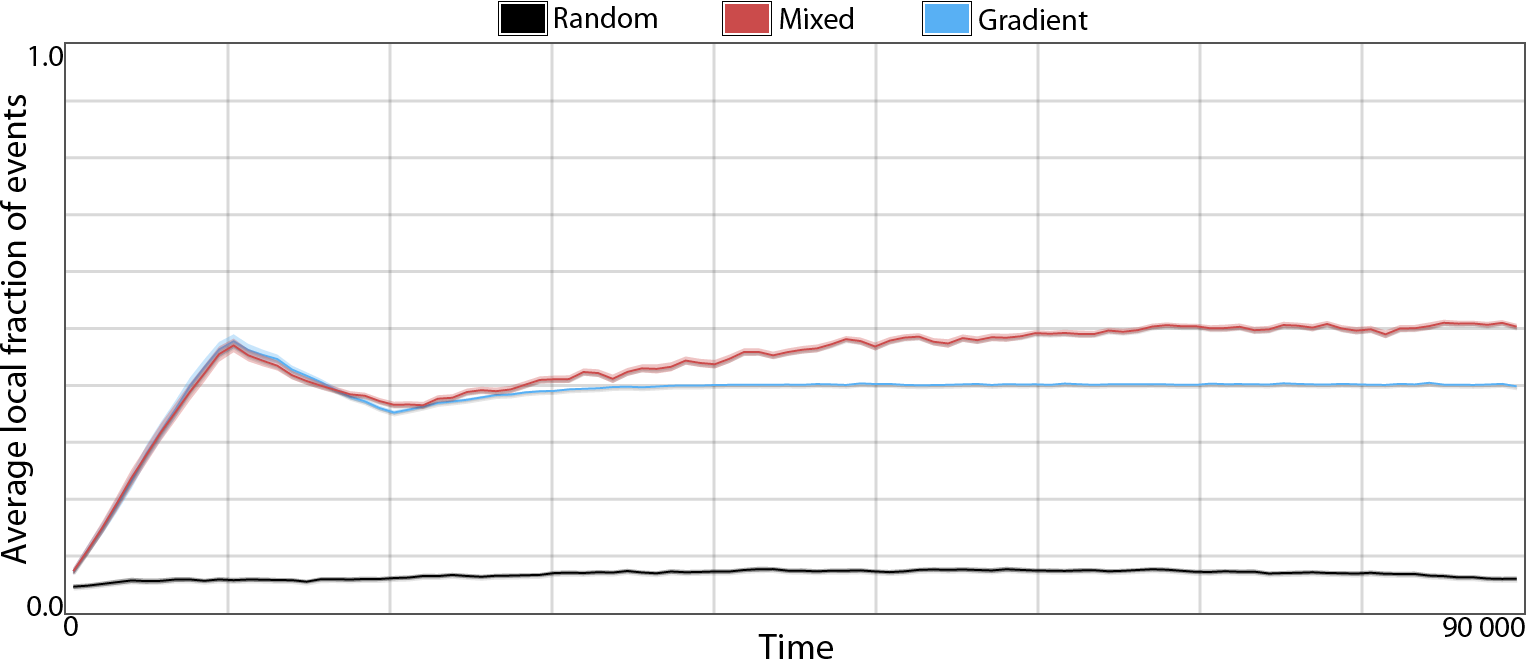}
	\caption{}
	\label{fig:B2b}
\end{subfigure}
	\caption{Evolution in time of the average global (a) and local (b)
window-based metrics for Experiment~2. Window size is $900$. Confidence
intervals correspond to the $95\%$ level.}
	\label{fig:B2}
\end{figure}

\begin{figure}[p]
	\centering
\begin{subfigure}[b]{0.95\textwidth}
	\includegraphics[width=\textwidth]{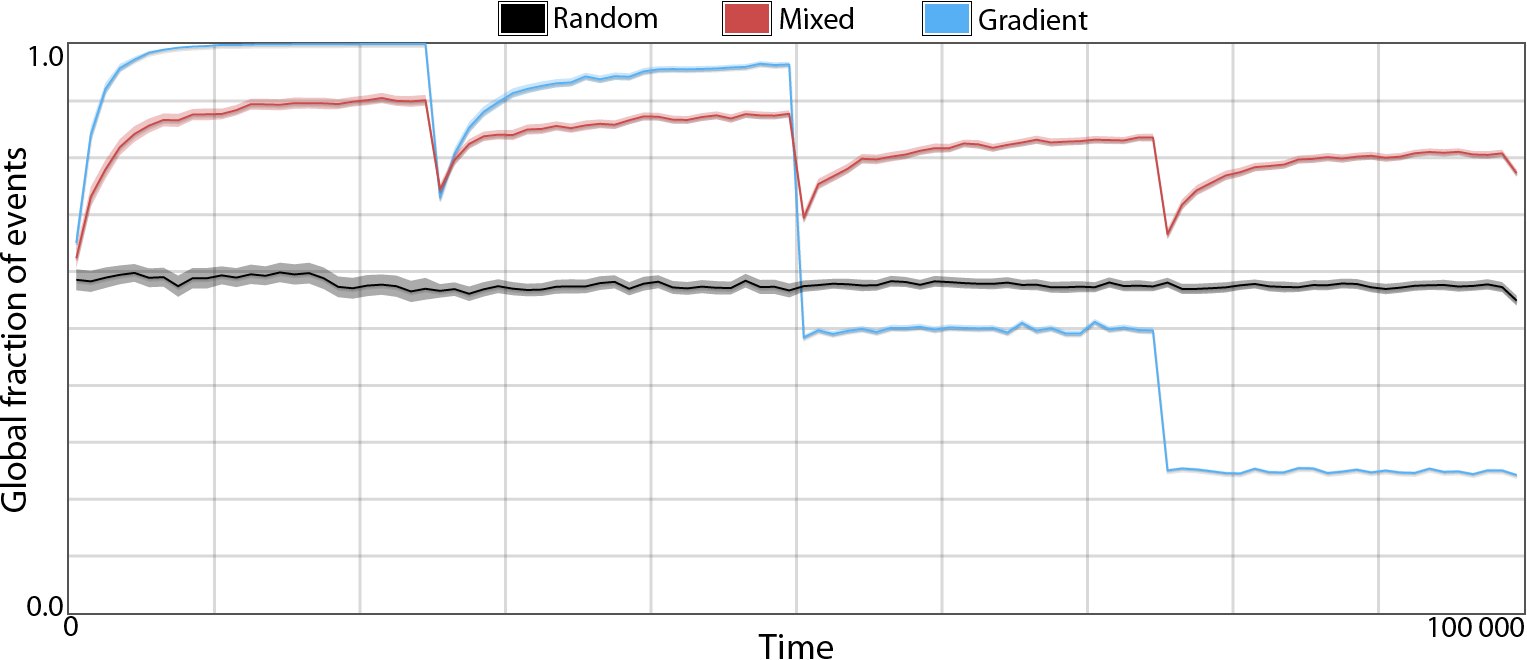}
	\caption{}
	\label{fig:B3a}
\end{subfigure}
\linebreak\linebreak
\begin{subfigure}[b]{0.95\textwidth}
	\includegraphics[width=\textwidth]{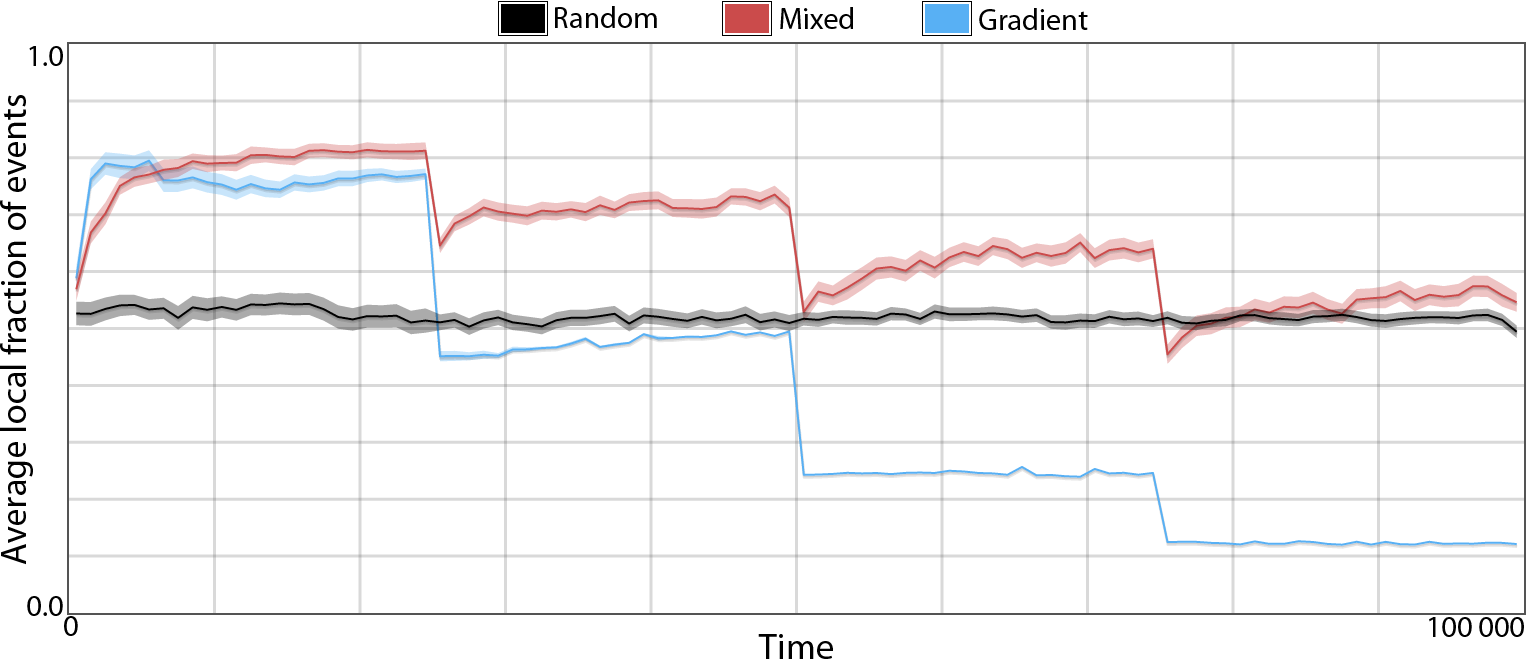}
	\caption{}
	\label{fig:B3b}
\end{subfigure}
	\caption{Evolution in time of the average global (a) and local (b)
window-based metrics for Experiment~3. Window size is $1\,000$. Confidence
intervals correspond to the $95\%$ level.}
	\label{fig:B3}
\end{figure}

\end{document}